\documentclass[%
 aip,
rsi,%
amsmath,amssymb,
 reprint,%
]{revtex4-1}

\usepackage{graphicx}
\usepackage{dcolumn}
\usepackage{bm}

\usepackage{epstopdf}

\usepackage{epstopdf}

\usepackage{dcolumn}

\usepackage{bbold}
\usepackage{gensymb}

\usepackage{hyperref}

\newcommand{\um}[1]{{\"#1}}

\newcommand{\T}{\ensuremath{{^\intercal}}}

\newcommand{\V}[1]{\ensuremath{\boldsymbol{#1}}}
\DeclareMathOperator*{\argmin}{argmin} 
\DeclareMathOperator{\sech}{sech} 
\newcommand{\Rom}[1]{%
  \textup{\uppercase\expandafter{\romannumeral#1}}%
}

\newlength{\figwidth}
\usepackage{color}

\begin{document}
\title{An extended transfer operator approach for time-consistent coherent set analysis}

\author{Benedict L\um{u}nsmann}
\affiliation{Max Planck Institute for the Physics of Complex Systems (MPIPKS), 01187 Dresden, Germany}

\author{Rahel Vortmeyer-Kley}
\affiliation{Leibniz Institute for Baltic Sea Research (IOW), 18119 Rostock-Warnem\um{u}nde, Germany}

\author{Holger Kantz}
\affiliation{Max Planck Institute for the Physics of Complex Systems (MPIPKS), 01187 Dresden, Germany}


\begin{abstract}
Coherent oceanic mesoscale structures, especially the non-filamenting cores of
oceanic eddies, have gained a lot of attention in recent years.
These Lagrangian structures are considered to play a significant role in oceanic
transport processes which, in turn, impact marine life, weather and potentially
even the climate itself.
Answering questions regarding these phenomena requires robust tools for the
detection and identification of these structures.
In this article, we use transfer operator ideas to develop a novel method for the
identification of weakly-mixing coherent volumes in oceanic velocity field data
sets.
Unlike other methods, the approach focuses on maximizing consistency over longer
time periods.
We employ a time-centralized transfer operator approach with practical
modifications to identify potential structures in predetermined domains and
couple adjacent time steps to decide how to conduct the final partitioning.
The analysis pipeline includes plausibility checks that give further insights
into the stability and coherence of the inferred structure.
The presented method is able to find changing masses of maximal coherence in
stationary and non-stationary toy models and yields good results when applied to
field data.
\end{abstract}

\maketitle

The following article has been submitted to \mbox{Chaos: An Interdisciplinary Journal
of Nonlinear Science}. After it is published, it will be found at
\href{https://publishing.aip.org/resources/librarians/products/journals/}{Link}.

\vspace*{1cm}

\begin{quotation}
  Eddies, non-filamenting coherent oceanic mesoscale structures, are considered
  to impact oceanic transport processes and marine life in many ways.
  Studying their impact necessitates the development of robust methods of
  identification.
  Here, we present and test an extended and modified two-step transfer operator
  approach that facilitates the extraction of time-consistent coherent sets.
\end{quotation}

\section{Introduction}
\label{sec:introduction}


Horizontal transport processes in the upper layer of the ocean are dominated by
hydrodynamic mesoscale structures like jets, fronts and eddies. 
Their emergence, disappearance and complicated interplay orchestrates an
ever-changing chaotic flow that stirs and mixes the involved fluid volume.
However, not all parts of the ocean surface mix equally fast.
Coherent volumes resist filamentation for finite-time. 
They coherently transport trapped water masses in ambient water of different
properties and thus contribute to the patchiness of scalars fields like
temperature and salinity \cite{Aristegui1997,Martin2003,Beal2011,Dong2014,Karstensen2015}.
This in return implies impacts on marine life
\cite{Martin2003,Sandulescu2007,DOvidio2010,Prants2012,Karstensen2015,McGillicuddy2016} and
possibly the climate \cite{Beal2011}.

Algae production, in particular, is affected by mesoscale structures in the velocity
field in various subtle and obvious ways.
These structures are responsible for the generation of variability and
filamentation of plankton patches on smaller scales
\cite{Bracco2000a,Martin2003} and appear to have a
strong impact on large scale plankton distributions presumably due to the
formation of hydrodynamic biological niches \cite{DOvidio2010}.

Eddies constitute one such potential niche. 
These mesoscale structures trap their rotating fluid volume while being able to
generate vertical currents whereby they actively change the biogeochemical
conditions for algae growth inside their boundaries
\cite{Gaube2014,Brannigan2016,McGillicuddy2016}.
In this regard, studies of toy models have shown that the restriction of
upwelling to the vicinity of eddy centers may result in overall reduced algae
production \cite{Martin2002} and entrainment of nutrients may result in a confined
bloom inside an eddy \cite{Sandulescu2007}.

In order to study the impact of eddies it is first
necessary to develop reliable methods for the boundary estimation of finite-time
coherent sets that constitute the eddy core.
This is a non-trivial task and especially difficult in turbulent coastal regions.

Two fundamentally different classes of such eddy boundary detection approaches
have to be distinguished: traditional Eulerian methods and more recent Lagrangian
methods.
Eulerian methods operate on velocity field snapshots.
Popular examples are the Okubo-Weiss criterion
\cite{Okubo1971,Weiss1991,Isern-Fontanet2003,Chaigneau2008} and
any SSH-field (sea surface height) \cite{Itoh2010,Chelton2011,Gaube2014},
streamline \cite{Nencioli2010} or vorticity \cite{Mcwilliams1990} based
approach.
Lagrangian methods focus on trajectories of fluid parcels.
This class includes FTLE/FSLE (finite time/size Lyapunov exponent)
\cite{Boffetta2001,Shadden2005} based
methods, Lagrangian descriptors\cite{Mancho2011,Vortmeyer-Kley2016}, simple
clustering approaches\cite{Hadjighasem2015,Froyland2015}, geometric
approaches \cite{Haller2013,Farazmand2014,Haller2015} and transfer operator
based approaches \cite{Froyland2009,Froyland2010,Froyland2013,Ma2013} (for
a comparison of methods see \cite{Hadjighasem2017}).
Since coherent volumes are of Lagrangian nature, only approaches of the latter
class are able to provide accurate results.
Yet, Eulerian methods are computationally less expensive and have proven to
yield good approximations in real velocity fields.

In this paper, we present a novel two-step approach based on transfer operators
for the inference of coherent eddy cores.
The approach requires a preselected sequence of regions that follows the
temporal development of a potentially coherent structure.
First, each region of interest is analyzed independently using a modified
time-centralized transfer operator approach to quantify the affiliation of
individual region parts to the central eddy core (compare \cite{Froyland2009}).
The modifications account for coastal boundary fluxes as well as domain
separation and enforce the inference of circular structures (compare
\cite{Lunsmann2018}).
In the second step, we use short-time transfer operators to couple adjacent
time-steps (compare \cite{Froyland2015a}).
This way, for given partitions, we are able to compute the probability to stay
within the boundaries of the estimated coherent core.
The partitionings are then optimized to maximize the overall probability to stay
within these boundaries.
At several points in the analysis pipeline, we check for changes in the
coherence to guarantee the plausibility of the returned solution.

Our approach has several advantages over existing alternatives. 
First, it focuses on temporal consistency and couples the results of individual
time steps to generate a reliable result over larger time windows.
Most other approaches yield structures which are instantaneous and localized
in time.
These methods construct results for other points in time simply by integrating
the obtained solution.
Secondly, it decouples the size of the overall analysis window and the
integration time needed to define coherence.
Thus, the approach accepts a slight exchange of fluid volume across the inferred
boundaries while strongly reducing the generation of filaments.
This feature appears to be useful for the study of eddy cores over longer time
periods.

We test our approach on stationary velocity fields \cite{Lunsmann2018} and a
commonly used Bickley jet model \cite{Rypina2007} before studying its
performance on actual oceanic flows in the Western Baltic Sea.
Our results show that the approach is well able to infer coherent sets in
stationary and time-dependent toy model cases (see Sec.~\ref{sec:gbm} and
Sec.~\ref{sec:jet}).
As expected, actual oceanic flows prove to be more challenging.
The results illustrate how to tackle potential difficulties and point towards
additional insights that can be obtained using the proposed approach (see
Sec.~\ref{sec:baltic}).

The presented approach is successfully applied in \cite{Vortmeyer-Kley2018} to
facilitate the study of plankton population dynamics in coherent water masses in
the Baltic Sea.
There, the identification of coherent sets directly follows ideas of this article (see
Sec.~\ref{sec:baltic}).

\section{Method}
\label{sec:method}

\begin{figure}
  \centering
  \includegraphics[scale=1]{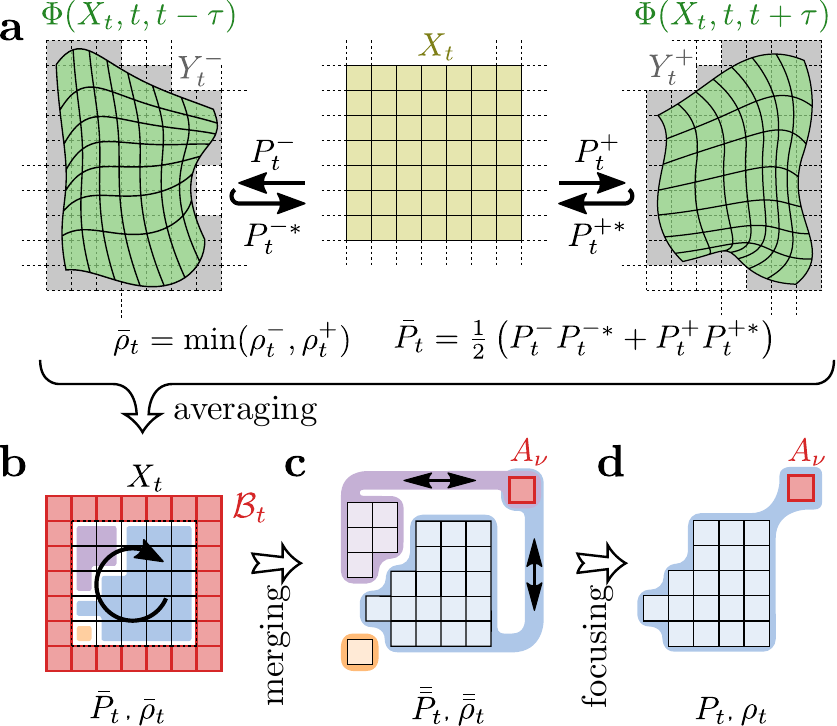}
  \caption{{\bf Construction of the modified time-centralized transfer operator $P_t$.}
    {\bf a)} For each domain $X_t$, we compute the transfer operators
    $P_t^+$, $P_t^-$ and their densities $\rho_t^+$, $\rho_t^-$ using the flow
    $\Phi$.
    Then, We compute the minimal mass vector $\bar{\rho}_t$ to determine the
    time-reversed operators $P_t^{+*}$ and $P_t^{-*}$.
    {\bf b)} The influences of future and past transport on mixing are averaged
    to create the time-centralized operator $\bar{P}_t$.
    We couple all potential filaments by merging the boundary $\mathcal{B}_t$
    (red) into one virtual tile $A_\nu$ (also red) and obtain
    {\bf c)} the transfer operator
    $\bar{\bar{P}}$.
    This graph of transfer probabilities potentially contains multiple
    components, some of which are coupled via the virtual tile $A_\nu$ (purple,
    blue) and some that might be disconnected (orange).
    {\bf d)} Focusing on the largest strongly connected component that is
    connected to the virtual tile $A_\nu$ (blue) yields $P_t$.
  }
  \label{fig:step1}
\end{figure}


The presented method is a modified time-centralized transfer operator approach
with two steps that aims to identify a coherent set contained in a
preselected sequence of regions.
In the following, we first explain step one, the individual treatment of regions
(Sec.~\ref{sec:toa}).
Subsequently, we show how adjacent time steps are coupled to optimize the partitioning in
step two (Sec.~\ref{sec:coupling}).
And finally, we introduce ways to check the plausibility of the returned
result (Sec.~\ref{sec:eval}).

\subsection{Transfer Operator Approach}
\label{sec:toa}

The presented approach aims, like most transfer operators, to capture the
material transport of a flow $\Phi$ by approximating its Frobenius-Perron
operator $\mathcal{P} $, an operator that describes how densities evolve
under $\Phi$.
Analyzing this operator allows the inference of sets that are almost-invariant
under evolution of the flow $\Phi$.

Since derivation of a closed mathematical solution of the Frobenius-Perron operator
$\mathcal{P}$ is rarely possible and thus unrealistic for actual oceanic flows,
the operator is approximated by a transfer probability matrix that specifies
the probability of transitions between different region parts under the
flow $\Phi$.
This graph of transfer probabilities can then be analyzed to find weakly
communicating partitions.
Essentially, the search for non-communicating sets becomes a graph cut problem.
In addition, the transfer probability matrix allows for modifications to focus
on specific structures \cite{Lunsmann2018} or to facilitate the analysis (see below).

We start with a sequence of regions that track and contain the trajectory of
exactly one eddy.
In the first step of the analysis, each region is treated individually using a
time-centralized transfer operator approach with certain modifications
(Sec.~\ref{sec:para_toa}).
The modifications include the treatment of coastal boundary fluxes
(Sec.~\ref{sec:para_coastal}), the
enforcement of circular solutions (Sec.~\ref{sec:para_mixing}) and measures
to avoid domain separation (Sec.~\ref{sec:para_lscc}).
The results of this analysis is a sequence of indicator vectors that quantify
the affiliation of region parts to the eddy core (Sec.~\ref{sec:para_indi}). 

\subsubsection{Transfer Operator}
\label{sec:para_toa}

In order to capture the transport properties of a time-dependent flow $\Phi:
\mathcal{X}\times\mathbb{R}\times\mathbb{R}\rightarrow\mathcal{X}$ at time $t$
over an integration time $\tau$, we introduce a domain-covering
partitioning of the domain of interest $X_t\subseteq\mathcal{X}$
\begin{align}
  X_t = \bigcup\limits_{i=1}^N A_{t,i} \quad\text{with}\quad A_{t,i}\cap A_{t,j}=\emptyset
 \;\forall i\neq j
\end{align}
and choose a partitioning that contains its image $Y_t^+:=\Phi(X_t,t,t+\tau)$ 
\begin{align}
  Y_t^+ \subseteq \bigcup\limits_{i=1}^M B_{t,i}^+ \quad\text{with}\quad B_{t,i}^t\cap B_{t,j}^t=\emptyset\;\forall i\neq j\;.
\end{align}
Furthermore, we define an appropriate mass vector $\rho_t\in\mathcal{R}_+^N$
that characterizes the mass contained in each subset, e.g., $\rho_{t,i} =
\mu(A_{t,i})$ with $\mu(\cdot)$ being the Lebesgue measure.
Then, we apply Ulam's method \cite{ulam1964problems} to approximate the
Frobenius-Perron operator
$\mathcal{P}(t,t+\tau)$ of the flow $\Phi(\cdot, t, t+\tau)$ (see
Fig.~\ref{fig:step1}a).

Accordingly, we approximate the transport probability $p(B_{t,j}, t+\tau |
A_{t,i}, t)$ from $A_{t,i}$ to $B_{t,j}$ by counting the fractions of tracers
that are injected in $A_{t,i}$ at time $t$ and arrive in $B_{t,j}$ at time
$t+\tau$:
\begin{align}
 p(B_{t,j}, t+\tau| A_{t,i}, t) \approx P^+_{t,ij} \\= \frac{\text{\# tracers that arrived in } B_{t,j}}{\text{\# tracers released inside } A_{t,i}}\label{eq:ulam}
\end{align}

Thus, the mass transported from tile $A_{t,i}$ to tile $B_{t,j}$ by the flow
$\Phi(\cdot, t, t+\tau)$ is given by $\rho_{t,i} P_{t, ij}^+$.

Analogously, we approximate the Frobenius-Perron operator
$\mathcal{P}(t,t-\tau)$ of the flow into the past $\Phi(\cdot, t, t-\tau)$ and
obtain another transport probability matrix $P_{t, ij}^-$ (see
Fig.~\ref{fig:step1}a).

\subsubsection{Reduction to Oceanic Transport}
\label{sec:para_coastal}

In oceanic flows the rules set for the flow across the coastline are crucial
for the investigation of coastal material transport.

For reasons of simplicity, we decided to ignore all mass that is washed ashore.
Hence, we focus on the water body that is only involved in pure oceanic
transport.
This is done by deleting all tracers that cross the coastline before computing
elements of the transfer probability matrix [Eq~(\ref{eq:ulam})] and adjusting
the entry of the mass vector $\rho_t$ in proportion of the number of deleted
tracers.

However, this way, the mass vector $\rho_t$ depends on the flow. Hence, we
generelly obtain two different mass vectors $\rho_t^+$ and $\rho_t^-$, one for
each time-direction.

The mass vector $\bar{\rho}_t$ that disregards all mass involved in coastal
boundary fluxes is thus given by
\begin{align}
  \bar{\rho}_{t,i} = \min(\rho_{t,i}^-, \rho_{t,i}^+)\;.
\end{align}

We use this mass vector $\bar{\rho}_t$ as a common basis for mass transport into
the future and past.

\subsubsection{Time-centralized transfer operator}

We then compute the time-centralized transfer probability matrices $\bar{P}_t$ using
the transfer operators in both time directions $P_t^\pm$ and their common mass
vector $\bar{\rho}_t$ (see Fig.~\ref{fig:step1}a).

This is done by using detailed balance to obtain the time-reversed operator of
$P_t^\pm$
\begin{align}
  P_{t,ij}^{\pm,*} := \frac{\bar{\rho}_j}{\sum\limits_j \bar{\rho}_j P_{t,ji}^\pm}P_{t,ji}^\pm
\end{align}
and by averaging future and past effects
\begin{align}
 \bar{P}_t:=\frac{1}{2}\left( P_t^+P_t^{+*} + P_t^-P_t^{-*} \right)\;.
\end{align}

The operator $\bar{P}_t$ describes the joint effects of transporting mass into
the future and past, adding diffusion (generated by coarse-graining) and
transporting back into the present \cite{Froyland2009}.

\subsubsection{Mixing Boundary}
\label{sec:para_mixing}

We plan to partition the graph of transfer probabilities described by the
time-centralized transfer operator $\bar{P}_t$ in subgraphs with minimal
inter-subgraph material transport such that one of the subgraphs can be
identified with the central eddy core.

Since, in most cases, the eddy is surrounded by multiple non-communicating
filaments, multiple cuts are necessary to uncover the central eddy core.
Many transfer operator approaches rely on clustering techniques to
determine the necessary number of cuts and subsequently the central structure of
the enclosed flow
\cite{Froyland2005,Froyland2007,speetjens2013footprints,Denner2015,Banisch2017}.
Instead, we argue, that it is simpler to reduce the number of necessary cuts
because the overall geometry of the eddy is known.
We know, that tiles at the boundary of the investigated domain are not part of
the eddy core but rather part of the filaments surrounding it.
By artificially connecting all filaments, we leave just one efficient graph-cut:
the boundary separating the inner eddy core from the outer embedding flow.
Essentially, this makes horizontal and vertical graph cuts inefficient and
enforces the inference of circular structures.

We thus merge all tiles contained in the boundary $A_{t,i}\in\mathcal{B}_t$ into
one virtual tile $A_{t,\nu}$ and modify the transfer probability matrix
$\bar{P}_t$ accordingly while leaving the structure of material transport inside
the boundary intact (see Fig.~\ref{fig:step1}b).
For simplicity, let $I=\{1,\ldots,N-n\}$ and $J=\{N-n+1, \ldots, N\}$ be the
indices of tiles outside and inside the boundary $\mathcal{B}_t$ and let
$\nu=N-n+1$ be the final index of the virtual tile.
Then the new transfer probability matrix $\bar{\bar{P}}_t\in\mathbb{R}_+^{\nu\times\nu}$ is given by
\begin{align}
  \bar{\bar{P}}_{t,ij} = \bar{P}_{t,ij}\;,\quad
  &\bar{\bar{P}}_{t,\nu j} = \frac{\sum\limits_{i\in J} \bar{\rho}_{t,i}\bar{P}_{t,ij}}{\sum\limits_{i\in J}\bar{\rho}_{t,i}} \;,\\
  \bar{\bar{P}}_{t,i\nu} = \sum\limits_{j \in J} \bar{P}_{t,ij}\;,\quad
  &\bar{\bar{P}}_{t,\nu\nu} = \frac{\sum\limits_{i\in J} \bar{\rho}_{t,i}\bar{P}_{t,ij}}{\sum\limits_{i\in J}\bar{\rho}_{t,i}}\;.
\end{align}
The new mass vector is
\begin{align}
  \bar{\bar{\rho}}_{t,i}= \bar{\rho}_{t,i}\;,\quad \bar{\bar{\rho}}_{t,\nu} = \sum\limits_{i\in J} \bar{\rho}_{t,i}\;.
\end{align}

\subsubsection{Largest Strongly Connected Component}
\label{sec:para_lscc}

The construction of the time-directed transfer operators $P_t^\pm$, the
restriction to oceanic fluid transport and the merging of the boundary may lead
to the separation of the domain in non-communicating regions.

Since we are not interested in the inference of isolated tiles (e.g., regions
that are cut-off by the cost-line), we reduce our analysis to the largest
strongly connected component that is connected to the boundary.
In other words, we drop anything that is isolated and if two or more components
are only connected via the virtual tile $A_\nu$ we keep the larger one (see Fig.~\ref{fig:step1}c/d).

The transfer probability matrix $\bar{\bar{P}}_{t}$ is modified to account for
these changes.
We refer to the resulting and final transfer operator as $P_t$ and the final
mass vector as $\rho_t$.

\subsubsection{Indicator Vector}
\label{sec:para_indi}

We seek to partition the graph described by the transfer operator $P_t$ into two
sets, inner core $C_t\subset X_t$ and outer flow $S_t\subset X_t$, such that the inter-set mass transport
$\mathcal{M}$ is minimized.
Let $z_t\in\{-1,1\}^\nu$ be an indicator vector such that $z_{t,i}=1\Leftrightarrow A_{t,i}\in
C_t$ then this mass transport is given by
  \begin{align}
    \mathcal{M} &= \frac{1}{4} \sum\limits_{i,j=1}^\nu \rho_{t,i} P_{t,ij}\left( z_{t,i}-z_{t,j} \right)^2\propto z_t\T L_t z_t\;,\\
    L_{t,ij} &=  \delta_{ij} \sum\limits_{k=1}^\nu\frac{\rho_{t,i}P_{t,ik}+\rho_{t,k}P_{t,ki}}{2} + \frac{\rho_{t,i}P_{t,ij}+\rho_{t,j}P_{t,ji}}{2}\;.
  \end{align}

Wherein $L_t$ is a Laplacian matrix and $\delta_{ij}$ is the Kronecker-delta.
We now search for an indicator vector $z_t$ that minimizes this expression.
Unfortunately, without further assumptions this problem is NP-hard.

Relaxing the problem by letting indicator elements take real values
$\tilde{z}_t\in \mathbb{R}^\nu$ while establishing conditions to avoid trivial
solutions ($\sum_i\tilde{z}_{t,i}\rho_{t,i} = 0$, $\sum_i\tilde{z}_{t,i}^2\rho_{t, i} = 1$)
eventually results in the minimization of Rayleigh quotient (compare \cite{Dellnitz1999})
\begin{align}
  \tilde{z}_t^* = \argmin_{\tilde{z}_t\;|\;\langle \tilde{z}_t, \rho_t \rangle=0} \frac{\tilde{z}_t\T L_t\tilde{z}_t}{\tilde{z}_t\T D_{\rho_t}\tilde{z}_t}\;,
\end{align}
where $D_{\rho_t}$ is a diagonal matrix with $\rho_t$ as its diagonal entries
and $\langle \cdot, \cdot \rangle$ is the standard scalar product.

The solution to this problem $\tilde{z}_t^*$ is given by the eigenvector to the second smallest
eigenvalue of $L_t$ \cite{Dellnitz1999,Lunsmann2018}.
We choose its sign such that the virtual tile $A_{t,\nu}$ has a negative entry.
Since the virtual tile is definitely not part of the inner set $C_t$, more positive
values indicate membership with the inner coherent core.
This eigenvector $\tilde{z}_t^*$ may be thresholded at different values to
generate an indicator vector $\bar{z}_t$ and partitions of different size.
In the next section, we present how to find the optimal threshold.

\subsection{Coupling of Adjacent Time Steps}
\label{sec:coupling}

\begin{figure}
  \centering
  \includegraphics[scale=1]{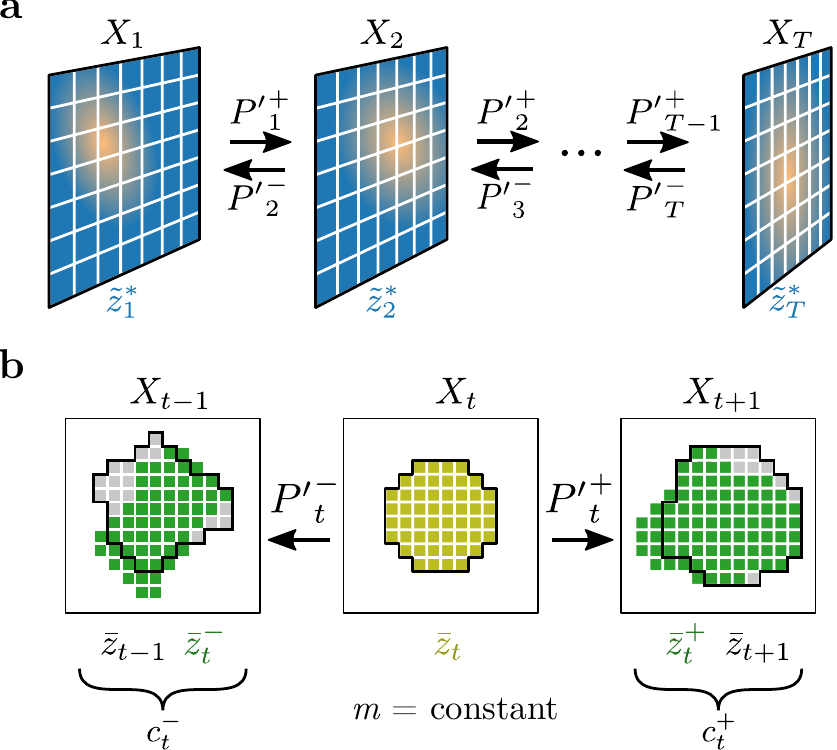}
  \caption{{\bf Generation of coherence ratios}
    {\bf a)} The analysis step is used to compute indicator vectors
    $\tilde{z}_t^*$ for each domain $X_t$ individually.
    In addition, we compute transfer operators ${P'}_t^\pm$ that connect
    adjacent time-steps.
    {\bf b)} We threshold each indicator vector such that the mass $m$ of the
    inner set is the same for each thresholded indicator vector $\bar{z}_t$.
    The vector $\bar{z}_t$ is transported via ${P'}_t^\pm$ and thesholded under
    the same premise to obtain $\bar{z}_t^\pm$. 
    Comparing the transported $\bar{z}_t^\pm$ with the thresholded indicator
    vectors of adjacent time-steps $\bar{z}_{t\pm 1}$ yields the coherence
    ratios $c_t^\pm$.
  }
  \label{fig:opt}
\end{figure}

At the end of the first analysis step, we obtained a sequence of independent real-valued
indicator vectors $\tilde{z}^*_t$ for each time-step $t$ that quantify the
affiliation of each tile $A_{t,i}$ with the coherent core $S_t\subset X_t$.
In the second step, we use thresholds for these indicator vectors
$\tilde{z}^*_t$ to partition the domains $X_t$ in inner coherent core $S_t$ and
outer surrounding flow $F_t$.
These thresholds are determined by coupling adjacent time-steps via short-time
transfer operators ${P'}_t^\pm$ (see Fig.~\ref{fig:opt}) and maximizing the
overall probability $c$ to stay within the inferred sequence of sets $S_t$.

For each individual time step $t$, we threshold the indicator vector
$\tilde{z}_t^*$ to obtain a thresholded indicator vector $\bar{z}_t$ that
determines the inner core $S_t$.
It is furthermore practical to define a thresholded indicator vector
$\bar{z}_t$ by the mass $m_t$ of the coherent core $S_t$ it defines.
Let $\Theta(\cdot)$ be the heavy-side function, then
\begin{align}
    \begin{split}
    \bar{z}_{t, i}(m_t) := \Theta\left( \tilde{z}_{t,i}^* - \vartheta_t(m_t) \right)\\
    \textrm{such that}\;\sum_i \bar{z}_{t,i}\rho_{i,t} \approx m_t\;
  \end{split}
\end{align}
defines the inner core $S_t$ with mass close to $m_t$ via
\begin{align}
  S_t(m_t) := \bigcup_{i\;|\;\bar{z}_{t,i}=1} A_{t,i}\;.
\end{align}
Computationally, the correct threshold $\vartheta_t$ for every mass $m_t$ can be
found using a line search.

Thus, given a sequence of masses $m_t$, we generate thresholded indicator
vectors $\bar{z}_t(m_t)$ that define a sequence of estimated coherent sets $S_t$
with approximately these masses.

However, these partitionings should not be independent from each other since
they all describe the evolution of the same coherent structure.
So, in order to couple domains $X_t$, $X_{t\pm 1}$ that are adjacent in time, we
compute short-time transfer operators ${P'}_t^\pm$ that quantify the transport
from one domain $X_t$ to the next $X_{t\pm 1}$ using Ulam's method (see
Fig.~\ref{fig:opt}a), one for each time direction.
From there, we compute normalized transfer operators $Q_t^\pm$ by means of
\begin{align}
  \eta_t^\pm &= \rho_{t}{P'}_t^\pm\;\textrm{and}\\
  Q_t^\pm &= D_{\rho_t} {P'}_t^\pm D_{\eta_t^\pm}^{-1} \;.
\end{align}
Here, $D_v$ is a diagonal matrix with $v$ on its diagonal.
These operators conserve the one-vector and can thus properly
transport the indicator vectors $\tilde{z}_t^*$ into the future and past
\begin{align}
  \tilde{z}_{t}^{*,\pm} =  \tilde{z}_{t}^* Q_t^\pm\;.
\end{align}
These indicator vectors $\tilde{z}_{t}^{*,\pm}$ are again thresholded such that
the resulting partitioning in the adjacent domains $X_{t\pm 1}$ have the mass
$m_t$, i.e.
\begin{align}
  \bar{z}_{t, i}^\pm &= \Theta(\tilde{z}_{t,i}^{*, \pm} - \vartheta_t^\pm)\;
                       \textrm{such that}\;
                       \sum_i\bar{z}_{t,i}^\pm\eta_{i,t} \approx m\;.
\end{align}
This defines the a transported coherent core $S_t^\pm$ in the adjoining domains
by
\begin{align}
  S_t^\pm(m_t) = \bigcup_{i\;|\;\bar{z}^\pm_{t,i}(m_t)=1} A^\pm_{t\pm 1,i}\;.
\end{align}

Now, the probability of starting in the set $S_t\subset X_t$ and being transported into the
next set $S_{t\pm 1} \subset X_{t\pm 1}$ can be expressed by
\begin{align}
  \begin{split}
  &p(x_{t\pm1}\in S_{t\pm1}| x_{t}\in S_{t}) \\
  &= 
  \frac{\mu\left(\Phi^{-1}(S_{t\pm 1},t\pm 1,t)\cap S_t\right)}{\mu(S_t)}\\
  &\approx
  \frac{\bar{z}_t^\pm D_{\eta_t^\pm} \bar{z}_{t\pm1}}{\bar{z}_t^\pm D_{\eta_t^\pm} \bar{z}_t^\pm}:=c_t^\pm\;.
  \end{split}
\end{align}
This defines $c_t^\pm$, the next-time coherence ratios of transport into the future
and past (see Fig.~\ref{fig:opt}b).
Averaging these coherence ratios geometrically yields 
\begin{align}
  c^\pm = \sqrt[T-1]{\prod_t c_t^\pm}\;,
\end{align}
average coherence ratios that depict the probability to stay
in the inferred inner cores $S_t$ for one time step in each time direction.
We define the average of these quantifies, the total averaged coherence ratio
\begin{align}
  c = \frac{1}{2}\left( c^+ + c^- \right)\;,
\end{align}
to be the target function of our optimization.

Maximizing the total averaged coherence ratio is no simple task since it
critically depends on all masses $m_t$.
Different options are conceivable to facilitate the optimization process.
For example, it would be possible to use a greedy or block-greedy combination of
line searches starting from the end points $t=1,T$. 
It would also be conceivable to fix a minimum value of either $c_t^+$ or $c_t^-$,
to maximize the other and to follow the corresponding time direction.
Both would lead to different results that focus on different aspects of coherent
sets.

Under the condition that the underlying flow $\Phi$ is not too divergent, we
here propose to simply set $m_t=m$ for all $t$ with the idea that the Lebesgue
mass of the coherent volume should not change significantly.
This reduces the optimization of $T$ masses to the question which mass $m$
maximizes the total averaged coherence ratio $c$.

\subsection{Evaluating Results}
\label{sec:eval}

Two different measures are useful to investigate the validity during analysis:
the sequence of second smallest eigenvalues $\lambda_t$ used to compute the
eigenvector $\tilde{z}^*_t$ and the sequences of coherence ratios $c_t^\pm$.

The larger $\lambda_t$ the less coherent the partitioning
determined by thresholding $\tilde{z}^*_t$ will be.
Hence, we will find peaks in the eigenvector $\lambda_t$ whenever the
structures in the flow become less coherent.

The sequence of coherence ratios $c_t^\pm$ for a given mass $m_t$ but most
likely for all masses will show a similar behavior.
However, since the coherence ratios operate on the thresholded indicator vectors
$\bar{z}_t$, the impact of sudden changes in the velocity field will be more
dramatic.
Rapid changes in $c_t^\pm$ also indicate moments in which the assumption that
the mass of the coherent volume does not change is not fulfilled.
In these moments filaments might be shed or entrained or the eddy might become
generally unstable and vanish in favor of another structure.

This means, that the investigation of the evolution of eigenvalues $\lambda_t$ and
the sequence of coherence ratios $c_t^\pm$ gives insights into the development
of the coherent water mass and helps to check the plausibility of the assumptions
that form the basis of the following optimization.

More practically, it helps to find the intervals in which the existence of
coherent volumes of constant mass is probable and helps to identify where the
method is not able to find concrete results.
We will see that this comes in handy if real oceanic data is investigated.

\section{Results}
\label{sec:results}

We test the presented approach by means of three different models.
First, we investigate the qualifications of this approach using a
stationary velocity field of Gaussian vortices (see \cite{Lunsmann2018}).
In the second test, we aim to find the coherent set in the wake of a
Bickley-jet \cite{Rypina2007}, a non-stationary flow and standard in the field
\cite{Hadjighasem2017}.
And finally, we apply our approach to real oceanic velocity fields of the Baltic
Sea.

In all cases we follow the same scheme and use an linearly interpolated gridded
velocity field in discrete time as starting point of our investigations since
this is the data format of real oceanic velocity fields.
Trajectories in this velocity field are generated by means of numerical integration
using Heun's method.

\subsection{Stationary Flow}
\label{sec:gbm}

First, we test our approach using a stationary two-dimensional velocity field.
Since, in these flows, separatrices form the natural barriers between coherent
sets of maximal size, we are able to investigate whether and to which
extent our approach is able to recover the ground truth in a simple scenario.

For this check, we use a stationary Gaussian blob model. 
Velocity fields generated by this approach are given by
\begin{align}
  \V{v}(\V{x}) = -\frac{1}{2\pi} \sum\limits_{i=1}^M
  \frac{\Gamma_i\left( 1-\exp\left[ -\frac{(\V{x}-\V{x}_i)^2}{2\sigma_i^2} \right] \right)}{(\V{x}-\V{x}_i)^2}
  (\V{x}-\V{x}_i)\times\hat{\V{e}}_z
\end{align}
where $\Gamma_i$ and $\sigma_i$ are vorticity and standard deviation of
individual Gaussian vortices.
Time and distances are measured in arbitrary units.

\begin{figure*}
  \centering
  \includegraphics[scale=1]{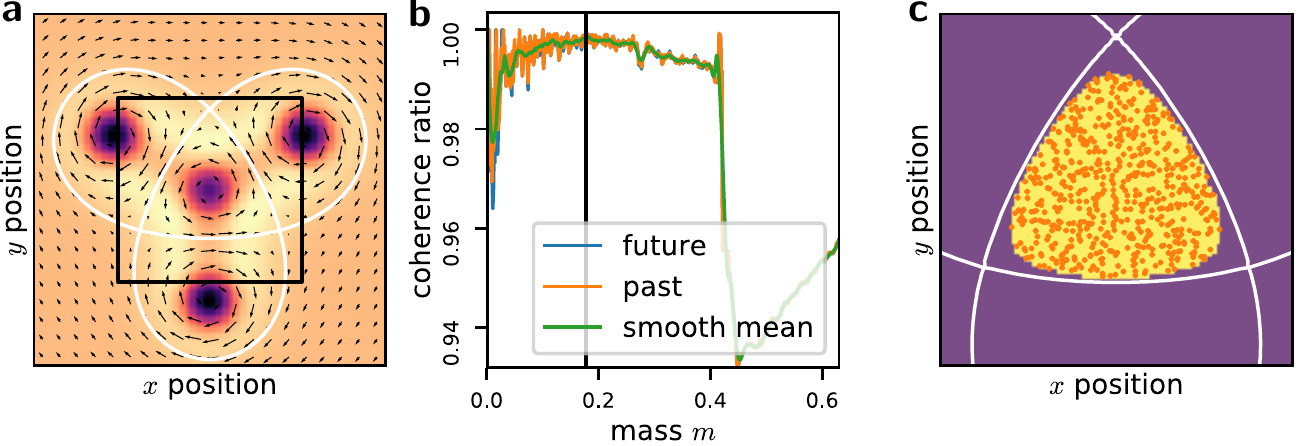}
  \caption{{\bf Stationary flow.}
    {\bf a)} Velocity field of flow with Okubo-Weiss criterion in the
    background. White: Separatrices. Black: Domain choice.
    {\bf b)} Average future $c^+$ (blue) and past $c^-$ (orange) coherence
    ratios display a rather noisy signal.
    Smoothing the total mean coherence ratio $c$ (green) results in a curve with
    a distinct maximum (black vertical line).
    {\bf c)} The partitioning defined by the mass of this maximum results in a
    good approximation of the inner set (yellow) that fills the separatrices.
    Tracers (orange) do rarely cross the boundary and stay mostly inside the
    inferred inner region.
  }
  \label{fig:gbm}
\end{figure*}

For our test, we choose the same parameters as in \cite{Lunsmann2018}, i.e.
three similar vortices with negative vorticity surrounding a vortex with
positive vorticity.
The aim is to recover the maximal central coherent set and thus the separatrices
of the central eddy (see Fig.~\ref{fig:gbm}a).

We use this model to generate a gridded velocity field in the quadratic
domain confined by the vertices $x = (\pm 2, \pm2)$ with a spatial resolution of
$\delta x = 0.008$.
Even though the model is stationary, we need a temporal resolution to treat the
model as a real data set.
Therefore, we choose a time step of $h=0.02$ between subsequent time steps and
create a sequence of velocity fields of length $T = 200$.

Using this data set, we first compute the Okubo-Weiss criterion $Q$.
On the basis of this field $Q$ we choose the domain of interest $X$ by hand (see
Fig.~\ref{fig:gbm}a).
Since the flow is stationary, we choose $X$ to be the same for each time step.

For the numeric integration of the velocity field, we use the minimal time step
of the data set $\delta t = 0.02$.
The integration time to generate the transfer operator is set to $\tau=0.9$ and
couple subsequent time steps with $\tau'=h=0.02$.
Next-time coherence ratios are computed for $M=500$ different masses.

Since the model and the sequence of domains is stationary and the mass is forced
to be stationary too, the sequences of future and past next-time coherence
ratios $c^+_t, c^-_t$ are also stationary.
The averaged future and past coherence ratios $c^+_t, c^-_t$ first increases
strongly with mass $m$ and display a noisy plateau for intermediate masses $m$
before decreasing rapidly (see Fig.~\ref{fig:gbm}b).
This is the expected behavior for stationary flows that exhibit a foliated
hierarchy of coherent structures:
Each orbit around the central elliptic fixed point confines a coherent set.
Since all coherent sets are in principle equivalent and differences in the
coherence ratio are only caused by the placement and resolution of the tiled
covering, the averaged coherences rarely exhibit a distinct maximum.

However, larger coherent masses should on average appear more coherent than
smaller masses because of the tiling's finite resolution.
Thus, in order to decide which mass to choose for the threshold, we smooth the total
averaged coherence ratio $c$ using a standard Savitzky-Golay filter of third order
with a window size of $21$ steps.
The smoothed total average coherence ratio $\tilde{c}$ shows a clear unimodal
structure with a distinct maximum suitable for threshold selection (see
Fig.~\ref{fig:gbm}b).

This threshold yields a partitioning that satisfactorily approximates the
separatrices of the flow (see Fig.~\ref{fig:gbm}b).
Test tracers released in the recovered eddy core do not leave the inner set (up
to tiling resolution) for $249$ integration steps, more than five times the
observation horizon of the analysis of each individual time step.

\subsection{Bickley Jet}
\label{sec:jet}

Here, we study the results of our approach using a time-dependent Bickley jet
flow \cite{Rypina2007}.
The model that describes an idealized stratospheric flow of two interacting
Rossby waves is given by the stream function
\begin{align}
  \begin{split}
  \Psi(x,y,t) = &-U_0L \tanh\left( \frac{y}{L} \right) \\
  &+ U_0L\sech^2 \left( \frac{y}{L}\right) A_2\cos\left( k_2[x-c_2 t] \right)\\
  &+ U_0L\sech^2 \left( \frac{y}{L}\right) A_3\cos\left( k_3[x-c_3 t] \right)\;.
  \end{split}
\end{align}
where $L=1700$~km is the characteristic length scale and $U_0=62.66$~m/s is
the characteristic velocity. $k_n = 2n/r_e$ are meridional wave numbers on a
sphere with the radius of the earth $r_e = 6371$~km at $60$\degree latitude.
We adopt all parameter values from \cite{Rypina2007}, i.e. $c_2/U_0=0.205$,
$c_3/U_0=0.461$ and $A_2=0.1$ and $A_3=0.3$.
The velocity field is by given by $\V{v}(x,y,t)=(\partial_y \Psi(x,yt),
-\partial_x\Psi(x,y,t))$.
This parameter set results in a quasiperiodic
stream function that generates a meandering jet surrounded by a zone of
Lagrangian chaos and several vortices that move with constant velocity.

For the data set, we choose a spatial and temporal resolution of $\delta x =
80$~km and $\delta t = \frac{1}{4}$~h as well as a duration of $T = 10$~days.
We choose a small time step to reduce errors of non-symmetric numeric
integration for an integration time of $6$~days.

\begin{figure*}
  \centering
  \includegraphics[scale=1]{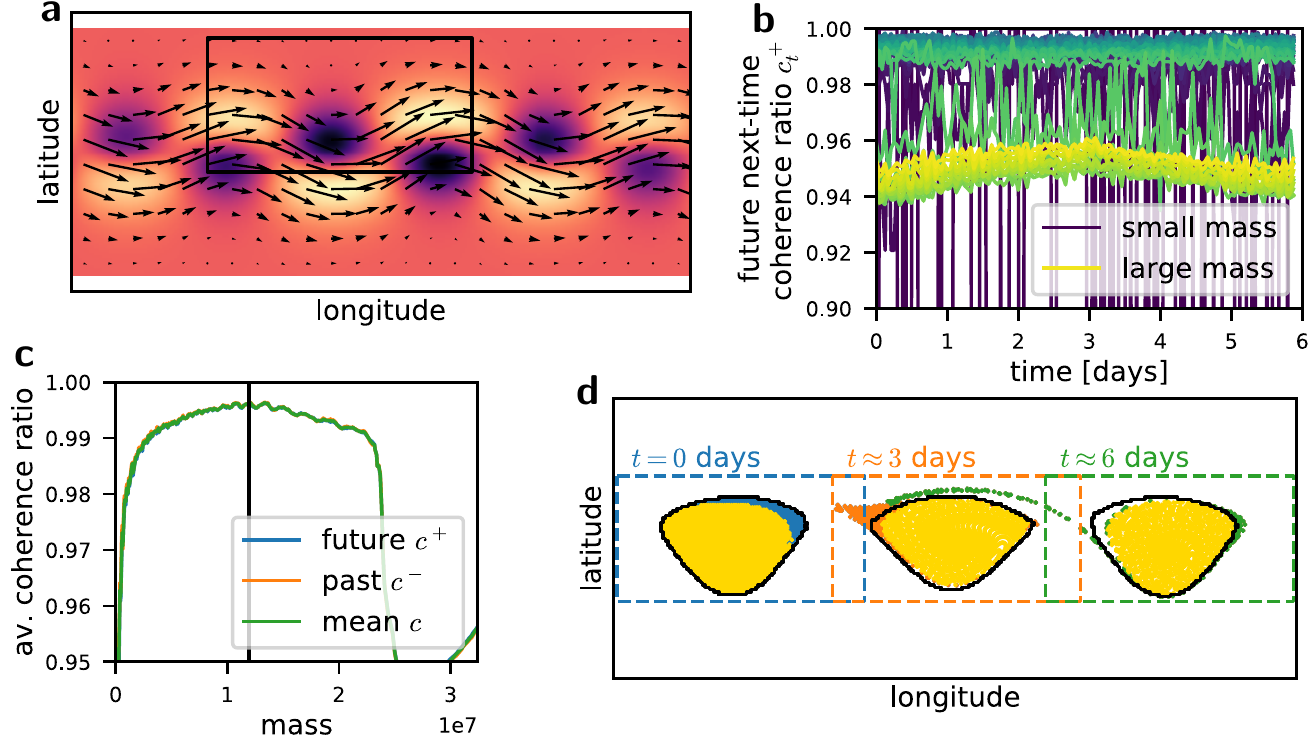}
  \caption{{\bf Analysis results for the Bickley jet.}
    {\bf (a)} Velocity field of the model with Okubo-Weiss field $Q$ in the
    background. A minimum of $Q$ is tracked and surrounded by a domain of
    interest $X$ (black).
    {\bf (b)} Future next-time coherence ratio $c^+_t$ for the complete analysis of
    $6$~days for $50$ of $500$ masses. Small masses (blue) result in
    inconsistent flickering and large masses (yellow) are less coherent than
    intermediate masses (light blue, green).
    {\bf (c)} Average future and past coherent ratios $c^+, c^-$ and total averaged
    coherence ratio $c$ dependent on the mass $m$.
    The figure shows a distinct maximum.
    {\bf (d)} Returned boundary estimates of the coherent volume (black) and
    considered domain of interest $X$ at three different times (blau, orange,
    green dashes).
    Comparison of integrated particles that are contained in the structure at
    all times (gold) and those that leave the structure at least once at
    different time (blue, orange, green dots).
    Only a few particles leave the structure and form filaments.
  }
  \label{fig:isf}
\end{figure*}

Using this data set we compute the Okubo-Weiss criterion and select an initial
domain of interest $X_1$ (see Fig.~\ref{fig:isf}a).
Domains for other times $X_t$ are selected by moving the initial domain $X_1$
with an appropriate constant velocity of $v_x=105$~km/h along the $x$-axis.
We set the integration time to $\tau=2$~days and couple intervals of
$\tau'=1$~h.
Each domain is partitioned into square tiles with a side length of $\Delta
x=80$~km.

Our approach generates future and past next-time coherence ratios $c^+_t,
c^-_t$ over $6$~days (see Fig.~\ref{fig:isf}b for $c^+_t$ as a reference).
While small sets apparently result in an inconsistent flickering of the
coherence ratio sequence, large sets are consistently less coherent than sets of
intermediate size.
The averaged future and past coherence ratios $c^+$ and $c^-$ display the same
weakly noisy and unimodal dependence on the mass $m$ that exhibits a distinct
maximum for intermediate masses (see Fig.~\ref{fig:isf}c).
Using the maximum of the total averaged coherence ratio $c$ as a threshold to
determine the best partitioning results in reasonable structures (see
Fig.~\ref{fig:isf}d).
The integration of test particles from the start of the analysis to the end of
the analysis reveals that most particles stay inside the structure, i.e. in all
inferred eddy cores (golden dots).
And not all test particles that leave the structure at some point (blue, orange,
green) do generate filaments; many remain in the vicinity of the uncovered eddy
core.

Parts of the leakage might be explained by ghosting, i.e. trajectories in the
vicinity of the actual eddy boundaries that diverge too slowly to be detected within
the considered observation horizon of $\tau=2$~days.

Apart from this small leakage at the corners, regions that are difficult so
resolve (see Fig.~\ref{fig:gbm}c for comparison), our method yields good
results.

\subsection{Baltic Sea}
\label{sec:baltic}

Finally, we apply our approach to a real data set of oceanic velocity fields.

The data set was generated by the coastal ocean model GETM (General Estuarine
Transport Model) \cite{Klingbeil2013,Klingbeil2018} of the Western Baltic Sea.
The setup of the model is chosen as in \cite{Grawe2015} and
\cite{Vortmeyer-Kley2016}.
The studied area has a horizontal spatial resolution
of $1/3$ nautical miles (approx. $600$~m). 
$50$ terrain-following adaptive layers focused towards stratification were used
for the vertical resolution.
In a post processing step the terrain-following coordinates were interpolated to
an equidistant vertical spacing of $1$~m and averaged over the upper $10$~m of
the water column to produce a quasi two-dimensional field.
The velocity fields are part of a multidecadal simulation and cover the timespan
March 2010 to October 2010. The temporal resolution of the velocity fields is
$1$~h.
More details of the coupled setup of GETM can be found in
\cite{Vortmeyer-Kley2016,Vortmeyer-Kley2018} where the data set was originally
used.

We use a Lagrangian descriptor, the MV-tool \cite{Vortmeyer-Kley2016}, to
identify all eddies with a lifetime longer than $100$~h
that travel more than $8$~km (see \cite{Vortmeyer-Kley2018} for details).
From this eddy data set, we select eight test eddies by hand for detailed
analysis.
Since the discussion of all eight eddies exceeds the scope of the
article, we focus on the analysis of eddy E1 and E2: Eddy E2
serves as an example of mostly effortless reconstruction while eddy E1
illustrates how the investigation of the future and past next-time coherence
ratio sequences help to find intervals of plausible coherence and improve our
results.

In all cases the sequence of domains of interest $X_t$ is generated
automatically on the basis of the eddy polygon returned by MV because the sheer
amount of data renders manual selection impractical (circa 250 time steps per
eddy).
For this purpose, we first analyze the distribution of polygon area provided by
the proxy and select an appropriate area value for all domains of interest.
Under the assumption that the mass of the coherent eddy core does not change,
the domain of interest should be larger than the size of the eddy core.
However, choosing an area value that is too large might incorporate additional
coherent volumes like slowly mixing filaments and other eddy cores that
interfere with the analysis.
Hence, an appropriate area value is much larger than the average polygon area
but smaller than any unreasonable outlier.
Next, we find the centroid of each polygon and its longitudinal and
latitudinal proportions.
The domain of interest in each time step is then chosen to be a rectangle with
the determined centroid, area and proportions.
This method of automatic domain selection compensates occasional rapid
changes of MV. 
However, also non-rectangular, automated domain selection methods are
conceivable.
In any case, minor changes in the geometry or the placement of the domain
should have no significant impact on the analysis.

Furthermore, we set the integration time for the transfer operator $P$ to
$\tau=36$~h and couple adjacent time steps, i.e. $\tau'_\textrm{E2}=1$~h and
$\tau'_\textrm{E1}=\{1,2\}$~h since some data was missing.
In order to enhance symmetry of numerical integration, we set the integration
time step to $\delta t=1/4$~h and interpolate linearly in time and space.

\begin{figure*}
  \centering
  \includegraphics[scale=1]{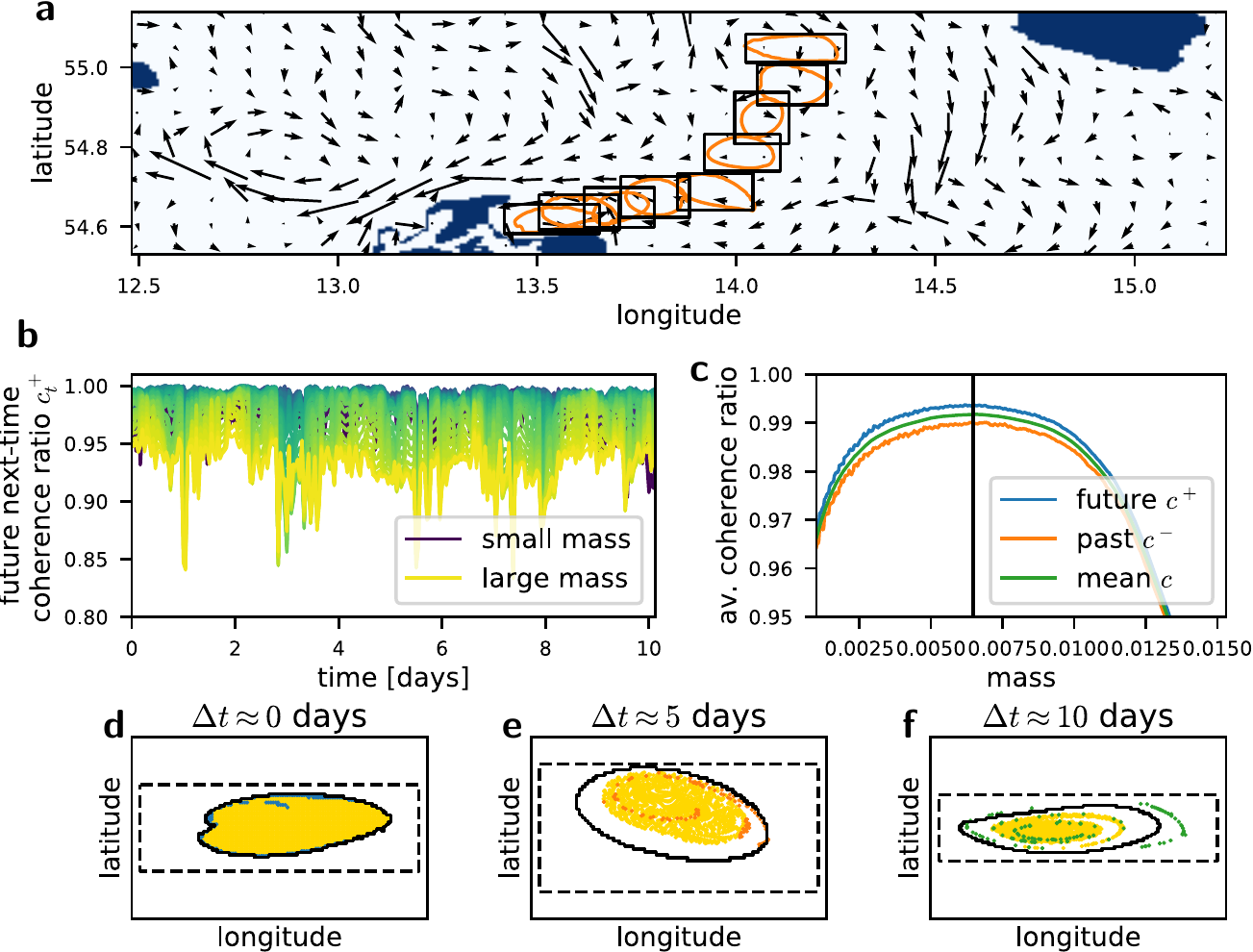}
  \caption{{\bf Eddy E2}
   {\bf (a)} Overview of eddy path and associated domains of interest. Every
   10th polygon is shown. Velocity field for $t=0$, i.e. 11:00 am, April 5th, 2010.
   {\bf (b)} The sequence of future next-time coherence ratios $c^+_t$ shows
   some structure but no rapid changes.
   Again intermediate masses show best results.
   Past next-time coherence ratios $c^-_t$ show a similar behavior.
   {\bf (c)} Averaged future and past coherence ratio $c^+, c^-$ and
   smoothed total mean coherence ratio $c$ depending on mass $m$ show distinct
   maximum used for thresholding (black).
   Persistent $c^+>c^-$ is an indicator for downwelling.
   {\bf (d/e/f)} Eddy boundary (black) and particle positions injected at the
   beginning at the beginning {\bf (a)}, middle {\bf (e)} and end {\bf (f)} of
   the structures life show that most particles remain within all detected
   boundaries (yellow).
   Particles that leave the boundaries at least once  mostly
   stay within the boundaries and produce only minor filaments
   (blue/orange/green).
   Contraction of particle cloud reveals downwelling.
  }
  \label{fig:eddy5675}
\end{figure*}

Eddy E2 starts from an upwelling zone at the coast of R\um{u}gen at April 5th,
2010 and travels north-east (see Fig.~\ref{fig:eddy5675}a).

Investigating the evolution of future and past next-time coherence ratios
$c^+_t, c^-_t$ reveals some but no significant structure: no drastic changes in
the coherence ratio are visible and intermediate masses yield the best results (see
Fig.~\ref{fig:eddy5675}b).

Averaging over time for each mass $m$ results in the averaged coherence ratio
curves $c^+, c^-$ (see Fig.~\ref{fig:eddy5675}c).
Smoothing the total averaged coherence ratio $c$ by means of an Savitzky-Golay
filter (window size $51$, order $3$) results in a distinct maximum that can be
used for thresholding.
Interestingly, the averaged future coherence ratio $c^+$ is persistently larger than
the averaged past coherence ratio $c^-$.
Since integration of particles is mostly time-symmetric within the observation
horizon, this is a strong indicator for a non-divergence free velocity field
with a sink: Mass is contracted and collected in the center of the structure and
thus reducing the probability to be transported across the boundary while the
opposite effect occurs in backwards time-direction.

Observing the trajectories of test tracers injected in the first inferred
boundary reveals that most tracers stay within \emph{all} following boundaries.
And even those particles which leave the structure at least once stay in the vicinity;
only few form filaments (see Fig.~\ref{fig:eddy5675}).
Moreover, the particles contract slightly what confirms a non-divergent free
velocity field presumably generated by downwelling.

\begin{figure*}
  \centering
  \includegraphics[scale=1]{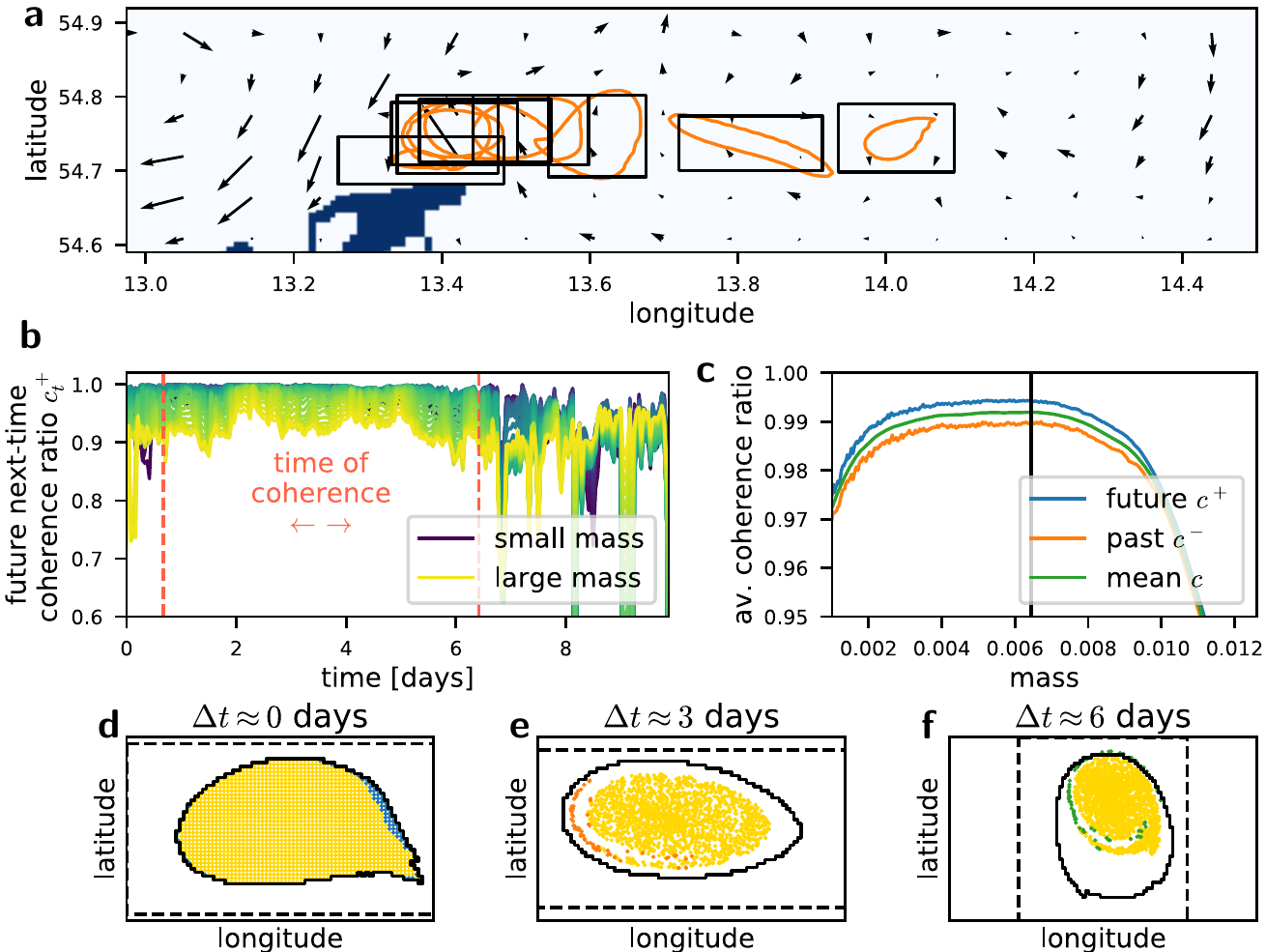}
  \caption{{\bf Eddy E1}
   {\bf (a)} Overview of eddy path and associated domains of interest. Every
   10th polygon is shown. Velocity field for $t=0$, i.e. 19:00 am, March 6th, 2010. .
   {\bf (b)} The sequence of future next-time coherence ratios $c^+_t$ shows
   rapid changes between which no real coherence exists.
   A window without rapid changes is identified for further analysis.
   {\bf (c)} Averaged future and past coherence ratio $c^+, c^-$ and
   smoothed total mean coherence ratio $c$ depending on mass $m$ display maximum
   for intermediate masses (black).
   Persistent $c^+>c^-$ is an indicator for downwelling.
   {\bf (d/e/f)} Again, eddy boundary (black) and particle positions injected at
   the beginning, at the beginning {\bf (a)}, middle {\bf (e)} and end {\bf (f)} of
   the structure's life show that most particles remain within all detected
   boundaries (yellow).
   Particles that leave the boundaries at least once  mostly
   stay within the boundaries and produce no minor filaments
   (blue/orange/green).
   Contraction of particle cloud reveals downwelling.
  }
  \label{fig:eddy1853}
\end{figure*}

Eddy E1 starts at the northern coast of R\um{u}gen at March 6th, 2010 where it
stays almost all its lifetime (see Fig.~\ref{fig:eddy1853}a).

Taking a look at the next-time coherence ratios reveals sudden drops and rapid
changes (see Fig.~\ref{fig:eddy1853}b).
We conclude that the presented method is not able to find a structure that is
coherent over the complete given time interval.
This might be caused by two factors: First, the assumptions necessary for the
application of the presented method are not fulfilled or, secondly, no eddy core
with stable mass exists over the full time interval.
The former occurs if the mass of the eddy core changes quickly (assumption of
constant mass) or domains where not chosen correctly (assumption of domain
consistency).
The latter occurs, if the eddy is deformed too much to justify coherence, e.g.,
when it collides with a front.
The structures detected by MV may loose their coherence by shedding
filaments. 
If a smaller coherent eddy core remains, old and new core might still be
detected as one consistent structure by MV.

MV is after all only a proxy that investigates time steps individually.
In any case, the analysis of next-time coherence ratios lets us identify time
intervals that are suited for the presented approach.
We simply choose a window of persistently high coherence to select a plausible
analysis window for the presented method.

Investigating the coherence ratios averaged over this time interval $c^+, c^-,
c$ ($c$ again smoothed using a filter Savitzky-Golay filter of third order
with a window size of $21$ steps) reveals a maximum that can be used to define
the threshold (see Fig.~\ref{fig:eddy1853}c).
We again observe consistently larger future next-time coherence ratios
than past next-time coherence ratios indicating a slight contraction of mass.

Investigating the trajectories of particles injected in the boundaries
at the beginning of the analysis interval, we find the same effects we already
found when evaluating the results of eddy E2:
Most particles stay in \emph{all} inferred boundaries, no large filaments are
generated and mass contraction is confirmed.

\section{Conclusion}
\label{sec:conclusion}

In this article, we presented a modified extended two-step transfer operator
approach following ideas of \cite{Froyland2009,Froyland2015a,Lunsmann2018}.

The first analysis step generates indicator vectors that provide a ranking of
domain parts for individual points in time.
Modifications of the transfer operator deal with coastline fluxes, help to focus
on larger water bodies and ensure that the domain is partitioned in an inner and
an outer set.
While the computation of indicator vectors follows ideas of classical transfer
operator methods \cite{Froyland2009} all modifications to the transfer probability matrix are
novelties in the style of \cite{Lunsmann2018} that increase overall performance.

In the second analysis step, we search for appropriate indicator vector
thresholds that result in one consistent and maximally coherent structure over time.
We assume a divergent free velocity field and relax the optimization procedure
to a line search.
During this step an appropriate time-interval may be chosen to guarantee that
critical assumptions hold.
Aside from general assumptions like persistent coherence and correct domain
selection that can be checked using the next-time coherence ratios,
it is also possible to check whether the velocity field is divergent
free, by looking at the averaged coherence ratios.
This analysis step constitutes a major change to former methods which mainly
focused on the treatment of isolated points in time.

We tested our approach using a stationary and a quasi-periodic model as well as
actual oceanic velocity field data.
In the stationary case, we were able to approximate the separatrices well (see
Fig.~\ref{fig:gbm}).
The uncovered structures stayed coherent for times much longer than the
observation horizon.
In the quasi-stationary case, a Bickley jet model, our approach resulted in a
sequence of boundaries that displayed high coherence.

Likewise, the study of real oceanic velocity fields yielded good results.
We analyzed eight Baltic eddies in total. 
The results are used in a study of plankton population dynamics in coherent eddy
cores \cite{Vortmeyer-Kley2018} to guarantee a common history of water parcels.

Here, two eddies have been discussed in detail.
The results of both eddies displayed high coherence; most particles stayed in
all inferred boundaries in forwards time direction (see Fig.~\ref{fig:eddy5675}
and Fig.~\ref{fig:eddy1853}.
Moreover, the application of our method to eddy E1 showed that the approach is able to
find plausible time windows for valid application (see Fig.~\ref{fig:eddy1853}b).
In both cases we noticed and confirmed that the velocity field was not
divergent free.
We found that mass was slightly contracting within the inferred boundaries and
thus that the assumption of volume conservation was not perfectly fulfilled.
Hence, particles injected into the last boundary and integrated backwards in
time leave the inferred structures with a significantly higher probability.

In summary, testing the presented approach was a success.
Our method found sequences of boundaries with high coherence in all scenarios.
In addition, our method is able to find appropriate time intervals for its
application which renders its results more trustworthy.
For future applications, further improvements are conceivable:
It is straight forward to improve the domain selection procedure by allowing
non-rectangular domain choices.
Instead of only coupling adjacent time steps, a weighted average over a range of
different time differences might also improve the approach. 
And most importantly, the usage of a more sophisticated optimization routines
would allow the treatment of divergent velocity fields.

In conclusion, we were able to show that modified transfer operator methods and
approaches that take temporal development of structures into account have high
potential of uncovering the boundaries of eddy cores.

\textit{Acknowledgments.}
We would like to thank Ulrike Feudel, Maximilian Berthold and Ulf Gr\um{a}we for
valuable discussions and helpful feed-back.

Moreover, we are grateful to Ulf Gr\um{a}we (Leibniz Institut for Baltic Sea
Research, Warnem\um{u}nde) for providing the velocity field
data set of the Baltic Sea and to the "`Norddeutsche Verbund f\um{u}r Hoch- und
H{\"o}chstleistungsrechnen (HLRN)"' for access to their high-performance
computing center. The analyzed model data was generated on the HLRN.

Most parts of the eddy tracking based on MV were performed at the HPC Cluster CARL, located
at the University of Oldenburg (Germany) and funded by the DFG through its Major
Research Instrumentation Programme (INST 184/157-1 FUGG) and the Ministry of
Science and Culture (MWK) of the Lower Saxony State.

Rahel Vortmeyer-Kley would like to thank BMBF-HyMeSimm FKZ 03F0747C for funding.

\end{document}